\documentclass{article}
\usepackage[latin9]{inputenc}
\usepackage{amsmath}
\usepackage{amssymb}

\makeatletter

\usepackage{amsfonts}
\usepackage{graphicx}
\providecommand{\U}[1]{\protect\rule{.1in}{.1in}}

\makeatother

\begin{document}
\begin{titlepage} \vspace{0.3cm}
 \vspace{1cm}
 
\begin{center}
\textsc{\Large{}{}{}{}{}{}{}{}{}{}\ \\[0pt] \vspace{0mm}
 Discrete Gravity }{\Large{}{}{}{}{}{}{}{}{}{} }\\[0pt]
\textsc{\Large{}{}{}{}{}{}{}{}{}{}\ \\[0.0pt] }{\Large\par}
\par\end{center}

\begin{center}
 
\par\end{center}


\begin{center}
\vspace{35pt}
 \textsc{ Ali H. Chamseddine$^{~a,b}$, Viatcheslav Mukhanov$^{~a}$}\\[15pt] 
\par\end{center}

\begin{center}
{$^{a}$ \textit{Ludwig Maximilian University, \\[0pt] Theresienstr.
37, 80333 Munich, Germany}}\\[0pt] {\small{}{}{}{}{}{}{}{}{}} 
\par\end{center}


\begin{center}
{$^{b}$ {\small{}{}{}{}{}{}{}{}{}{}}\textit{Physics Department,
}\\
 \textit{ American University of Beirut, Lebanon}}\\
 
\par\end{center}

\vspace{1cm}
 
\begin{abstract}
We assume that the points in volumes smaller than an elementary volume
(which may have a Planck size) are indistinguishable in any physical
experiment. This naturally leads to a picture of a discrete space
with a finite number of degrees of freedom per elementary volume.
In such discrete spaces, each elementary cell is completely characterized
by displacement operators connecting a cell to the neighboring cells
and by the spin connection. We define the torsion and curvature of the
discrete spaces and show that in the limiting case of vanishing elementary
volume the standard results for the continuous curved differentiable
manifolds are completely reproduced. 
\end{abstract}
\end{titlepage}

\section{Introduction}

One expects that scales smaller than the Planck length should not
make too much sense. In fact, for example, the perturbative quantization
of gravitational waves in Einstein theory leads to a  quantum metric
fluctuations of order one at these scales, suggesting that at sub-Planckian
scales the notion of a classical background manifold makes no sense.
For example, to ``test''\ such scales, we need to collide particles
with  center-of-mass energy exceeding the Planck energy, which by
the naive estimate should result in a black hole with a radius larger
than the Planck size. Therefore, it is natural to assume that the
``points''\ within the elementary Planck volume are indistinguishable
in any physical gedanken experiment, and instead of the differentiable
manifold to consider the space consisting of elementary cells with,
i.e. Planck volumes.

One might naturally expect that in such a case the more fundamental
theory must be similar to lattice gauge theory \cite{KWilson} or
the Regge simplex calculus \cite{Regge}. Lattice gauge theory is
based on the use of the elements of the Lie group (unitary matrices)
corresponding to the local group of gauge transformations. They are
defined on the vertices of the lattice constructed with Cartesian
coordinates in Euclidean space. It is not even clear how to formulate
this theory using, for instance, angular coordinates, let alone the
case of curved space. In the Regge calculus, on the other hand, the
basic building blocks are polyhedrons whose sides represent metric
variables and vertices represent curvature. Quite apart from the fact
that there is no clear unique way to get in the limit, a continuous
manifold, this construction presupposes the existence of  geometry
between two nearby vertices whose universal distance cannot be specified.
Therefore, this idea clearly contradicts the idea of universal structureless
building blocks of the elementary volume.

In this paper we develop the theory of discrete manifold consisting
of elementary cells, each characterized by its own tangent space and
displacement operators that allow us to move from a given cell to
the next one that has a common boundary with that cell. As a guiding
principle we use gauge symmetry of the local rotation group (associated
with each cell) and the elements of the spin connection group (see,
e.g., \cite{TKibble}) to form the corresponding curvature and other
gauge invariants. In a discrete space consisting of elementary cells,
each cell can be numbered by a series of integers which, in the continuous
limit, become coordinates on the corresponding differentiable manifold.
As we shall see, freedom in the choice of cells and the corresponding
shift operators lead to freedom in the choice of coordinates in the
continuous limit, respecting the diffeomorphism invariance of the
differentiable manifold. This is somehow reminiscent of the freedom
in the choice of conjugate variables in quantum theory, where the
corresponding conjugate operators satisfying the Heisenberg uncertainty
principle are related by canonical transformations. Regardless of
the choice of these variables, the uncertainty principle leads to
indistinguishability of states within a unit cell that were distinct
in classical phase space.

Finally, we will show that our formulas for discrete spaces reproduce
the standard formulas of differential geometry when the size of the
cells shrinks to zero. For simplicity, we consider only $d$-dimensional
space of Euclidean signature, where $d$ is arbitrary.

\section{Discrete cells manifold}

Let us consider a manifold consisting of cells of a given elementary
volume, e.g. Planck's volume, which we set as equal to one. Assuming
that points within each cell are physically indistinguishable, we
must first define what we mean by the dimension of the manifold consisting
of cells. In the continuous case, the $d$-dimensional manifold is
defined as a topological space for which each point has a neighborhood
homeomorphic to the Euclidean space $\mathbb{R}^{d}$. In our case,
we define the dimension $d$ assuming that each cell has $2d$ neighboring
cells that share a common boundary with each individual cell. Then
we can enumerate the elementary cells in the\textit{ d}-dimensional
space by a set of (positive and negative) $d$ integers 
\begin{equation}
\mathbf{n\equiv}\left(n^{1},n^{2},\cdots,n^{d}\right)\equiv n^{\mathbf{\alpha}},\label{eq:1}
\end{equation}
such that the points in neighboring cells that have a common boundary
are numbered such that only one of them differs by one unit. Next,
we define scalar functions $f\left(n^{\alpha}\right)$ that assign
only one number to each cell. This is the implementation of the idea
of assigning a finite number of degrees of freedom to each elementary
volume. In the case of the scalar field, this is one degree of freedom,
whereas for the massless vector field, for example, we have two degrees
of freedom per cell. Generalizing the definition of vectors to the
case of the discrete manifold, we have in each cell a set of displacement
operators $\mathbf{E_{\beta}}$ defined as 
\begin{equation}
\mathbf{E_{\beta}\mathnormal{(n)f(n^{\alpha})}\equiv\mathnormal{f(n^{\alpha}+\delta_{\beta}^{\alpha})}}\label{eq:2}
\end{equation}
where $\delta_{\beta}^{\alpha}$ is a Kronecker symbol. These operators
move us forward in $\alpha$-directions in the neighboring cells.
The $d$ operators $\mathbf{E_{\beta}}$ form a basis in a linear
$d$-dimensional space in each elementary volume. The sum and multiplication
by the real numbers $a,b$ in this space are defined by 
\begin{equation}
(a\mathbf{E}_{\beta}+b\mathbf{E}_{\gamma})f\left(n^{\alpha}\right)=af\left(n^{\alpha}+\delta_{\beta}^{\alpha}\right)+bf\left(n^{\alpha}+\delta_{\gamma}^{\alpha}\right).\label{eq:3}
\end{equation}
One can also introduce the inverse shift operators as 
\begin{equation}
\mathbf{E}_{\beta}^{-1}\left(n\right)f\left(n\right)\equiv f\left(n-1_{\beta}\right),\label{eq:3a}
\end{equation}
where, to simplify notation, we use (not bold) $n$ for $(n^{1},...n^{d})$
and $1_{\beta}$ shows which of $d$ arguments in $n$ is shifted
by unity. The shift operator in the cell is defined only when applied
to functions and operators in the same cell. Therefore, we must be
careful \textit{by applying first the left-most shift operator}, such
as in the following example, 
\begin{equation}
\mathbf{E}_{\alpha}(n)\mathbf{E}_{\beta}^{-1}(n)f(n)=\mathbf{E}_{\beta}^{-1}(n+1_{\alpha})f(n+1_{\alpha})=f(n+1_{\alpha}-1_{\beta}).\label{eq:4a}
\end{equation}
If we set $\alpha=\beta$, it follows from here that 
\begin{equation}
\mathbf{E}_{\alpha}(n)\mathbf{E}_{\alpha}^{-1}(n)=\mathbf{E}_{\alpha}^{-1}(n)\mathbf{E}_{\alpha}(n)=1,\label{eq:4b}
\end{equation}
i.e., the operators defined in (\ref{eq:2}) and (\ref{eq:3a}) are
indeed inverses of each other. Next, we define $d$ tangent operators
as 
\begin{equation}
\mathbf{e}_{\alpha}(n)\equiv\frac{1}{2}\left(\mathbf{E}_{\alpha}(n)-\mathbf{E}_{\alpha}^{-1}(n)\right).\label{eq:4c}
\end{equation}
As we shall see, in the continuous limit they become the tangent vectors
to the coordinate lines.

Assuming the existence of Euclidean scalar product in the linear space
of shift operators, we choose in each cell the \textit{d} orthonormal
operators $\mathbf{v_{\mathbf{\mathbf{a}}}}$, distinguished by Latin
indices $a,b,...=1,2,...d$, satisfying 
\begin{equation}
\mathbf{v}_{a}\bullet\mathbf{v}_{b}=\delta_{ab},\label{eq:5}
\end{equation}
and being vielbeins in each elementary volume. In a discrete space,
the local symmetry group with respect to which the theory must be
invariant is the group of rotations, under which the vielbein $\mathbf{v}_{a}(n)$
go to 
\begin{equation}
\widetilde{\mathbf{v}}_{a}(n)=\mathcal{R}_{a}^{b}(n)\mathbf{v}_{b}(n),\label{eq:26-1}
\end{equation}
where $\mathcal{R}_{a}^{b}(n)$ are the elements of the $SO(n)$ rotation
group in the vector representation. The rotations preserve the scalar
products (\ref{eq:5}), so that, $\widetilde{\mathbf{v}}_{a}\bullet\widetilde{\mathbf{v}}_{b}=\delta_{ab}$.
The operators $\mathbf{v}_{a}(n)$ can be expressed as linear combinations
of the tangent operators 
\begin{equation}
\mathbf{v}_{a}(n)=\bar{e}_{a}^{\alpha}(n)\mathbf{e}_{\alpha}(n),\label{eq:6}
\end{equation}
where $\bar{e}_{a}^{\alpha}(n)$ is the soldering form. Conversely,
$\mathbf{e}_{\alpha}(n)$ can be expressed in terms of the vielbein
$\mathbf{v}_{b}(n)$ as 
\begin{equation}
\mathbf{e}_{\alpha}(n)=e_{\alpha}^{b}(n)\mathbf{v}_{b}(n),\label{eq:7}
\end{equation}
where $e_{\alpha}^{b}(n)$ is inverse to $\bar{e}_{a}^{\alpha}(n)$,
that is, $e_{\alpha}^{b}\bar{e}_{a}^{\alpha}=\delta_{a}^{b}$. Note
that we will use $\bar{e}$ and $e$ for the soldering form and its
inverse to distinguish them when we need to deal with the individual
components, e.g., $\bar{e}_{1}^{1}$ from $e_{1}^{1}$. Here and later,
we adopt Einstein's summation convention for Greek and Latin repeated
indices \textit{only when they come in up and down positions}. Finally,
we can define the metric in each cell 
\begin{equation}
g_{\alpha\beta}(n)\equiv\mathbf{e}_{\alpha}(n)\bullet\mathbf{e}_{\beta}(n)=e_{\alpha}^{a}(n)e_{\beta}^{b}(n)\delta_{ab},\label{eq:7a}
\end{equation}
which gives meaning to the notion of volume of the cell.

\section{Parallel transport}

Next, we establish the rules for the parallel transport of the shift
operators and vielbein in discrete space. Let us consider the shift
from the cell $n+1_{\beta}$ to the $n$-cell. Then the corresponding
parallely transported tangent operator is defined as 
\begin{equation}
\mathbf{e}_{\alpha}^{p.t.}(n+1_{\beta}\rightarrow n)=\mathbf{e}_{\alpha}(n)+\Gamma_{\alpha\beta}^{\gamma}(n)\mathbf{e}_{\gamma}(n).\label{eq:8}
\end{equation}
For the vielbein we have to use the elements of the local spin connection
\textit{group}, so that 
\begin{equation}
\mathbf{v}_{a}^{p.t.}(n+1_{\beta}\rightarrow n)=(\varOmega_{\beta}^{-1}(n))_{a}^{b}\mathbf{v}_{b}(n),\label{eq:9}
\end{equation}
where $(\varOmega_{\beta}^{-1}(n))_{a}^{b}$ is inverse of 
\begin{equation}
\varOmega{}_{\beta}(n)=\exp\left(\omega_{\beta}^{cd}(n)J_{cd}\right).\label{eq:10}
\end{equation}
In the continuous limit, $\omega_{\beta}^{cd}$ is the well-known
spin connection. Here, $J_{cd}$ are the generators of the rotation
group, satisfying the following commutation relations 
\begin{equation}
\left[J_{ab},J_{cd}\right]=\frac{1}{2}\left(\delta_{bc}J_{ad}+\delta_{ad}J_{bc}-\delta_{ac}J_{bd}-\delta_{bd}J_{ac}\right),\label{eq:10a}
\end{equation}
which in equation (\ref{eq:9}) must be taken in the vector representation
\begin{equation}
(J_{cd})_{\;a}^{b}=\frac{1}{2}(\delta_{c}^{b}\delta_{da}-\delta_{ac}\delta_{d}^{b}).\label{eq:10b}
\end{equation}
Accordingly, both the affine and spin connections, $\Gamma$ and $\omega$,
can be fully expressed in terms of the soldering forms and their differences
in the neighboring cells. To obtain the corresponding equation for
them, we first note that it follows from (\ref{eq:7}) 
\begin{equation}
\mathbf{v}_{a}(n)\bullet\mathbf{e}_{\alpha}(n)=e_{a\alpha}(n),\label{eq:11}
\end{equation}
where the Latin indices are moved up and down with the Kronecker symbol
$\delta_{b}^{a}\equiv\delta_{ab}$. In the parallel transport the
scalar product does not change and thus on the one hand 
\begin{equation}
\mathbf{v}_{a}^{p.t.}(n+1_{\beta}\rightarrow n)\bullet\mathbf{e}_{\alpha}^{p.t.}(n+1_{\beta}\rightarrow n)=\mathbf{v}_{a}(n+1_{\beta})\bullet\mathbf{e}_{\alpha}(n+1_{\beta})=e_{a\alpha}(n+1_{\beta}),\label{eq:12}
\end{equation}
while on the other hand, using the definitions in (\ref{eq:8}) and
(\ref{eq:9}) we obtain 
\begin{equation}
\mathbf{v}_{a}^{p.t.}(n+1_{\beta}\rightarrow n)\bullet\mathbf{e}_{\alpha}^{p.t.}(n+1_{\beta}\rightarrow n)=(\varOmega_{\beta}^{-1}(n))_{a}^{b}e_{b\alpha}(n)+(\varOmega_{\beta}^{-1}(n))_{a}^{b}\Gamma_{\alpha\beta}^{\gamma}(n)e_{b\gamma}(n).\label{eq:13}
\end{equation}
From here we conclude that 
\begin{equation}
e_{a\alpha}(n+1_{\beta})=(\varOmega{}_{\beta}^{-1}(n))_{a}^{b}e_{b\alpha}(n)+(\varOmega_{\beta}^{-1}(n))_{a}^{b}\Gamma_{\alpha\beta}^{\gamma}(n)e_{b\gamma}(n).\label{eq:14}
\end{equation}
Multiplying both sides by $\varOmega$ we get the expression for the
affine connection 
\begin{equation}
\Gamma_{\alpha\beta}^{\gamma}(n)e_{a\gamma}(n)=(\varOmega_{\beta}(n))_{a}^{b}e_{b\alpha}(n+1_{\beta})-e_{a\alpha}(n),\label{eq:15}
\end{equation}
which allows us to express the affine connections in terms of spin
connection and soldering form.

\section{Torsion and Curvature}

Assuming that there is no torsion, i.e. $\Gamma_{\alpha\beta}^{\gamma}=\Gamma_{\beta\alpha}^{\gamma}$,
from (\ref{eq:15}) we find that the condition of no-torsion takes
the following form 
\begin{equation}
(\varOmega_{\beta}(n))_{a}^{b}e_{b\alpha}(n+1_{\beta})-e_{a\alpha}(n)=(\varOmega_{\alpha}(n))_{a}^{b}e_{b\beta}(n+1_{\alpha})-e_{a\beta}(n),\label{eq:17}
\end{equation}
These equations can be solved, to express $\omega_{\beta}^{cd}(n)$
entirely in terms of the soldering forms $e_{a\beta}$ in cell $n$
and the neighboring cells $n+1$. To see how this can be done explicitly,
it is more convenient to use the spinor rather than the vector representation,
in which the generators of the rotation group in (\ref{eq:10a}) become
\begin{equation}
J_{ab}=\frac{1}{8}(\gamma_{a}\gamma_{b}-\gamma_{b}\gamma_{a}),\label{eq:18}
\end{equation}
where the Dirac gamma matrices are assumed to be Hermitian and satisfy
the Clifford algebra 
\begin{equation}
\left\{ \gamma_{a},\gamma_{b}\right\} \equiv\gamma_{a}\gamma_{b}+\gamma_{b}\gamma_{a}=2\delta_{ab}.\label{eq:19}
\end{equation}
By incorporating vielbeins $e_{\alpha}^{a}$ into the Clifford algebra,
i.e., by introducing the matrices 
\begin{equation}
e_{\alpha}(n)\equiv e_{\alpha}^{a}\left(n\right)\gamma_{a},\label{eq:20}
\end{equation}
we can use the standard methods (see, e.g., \cite{CW}) to convert
the no-torsion condition (\ref{eq:17}) into the following useful
form 
\begin{equation}
(\varUpsilon_{\beta}(n)e_{\alpha}(n)\varUpsilon_{\beta}^{-1}(n)-e_{\alpha}(n))-(\alpha\leftrightarrow\beta)=0,\label{eq:21}
\end{equation}
where 
\begin{equation}
\varUpsilon{}_{\alpha}(n)\equiv\varOmega_{\alpha}(n)E_{\alpha}(n),\label{eq:22}
\end{equation}
and $\varOmega_{\alpha}(n)$ are given in (\ref{eq:10}), while $J$
are defined in (\ref{eq:18}). From now on and in the future we will
use non-bold notation for the shift operators to avoid cumbersome
looking formulas. Taking into account that 
\begin{equation}
e^{\omega}e_{\nu}e^{-\omega}=e_{\nu}+\left[\omega,e_{\nu}\right]+\frac{1}{2!}\left[\omega,\left[\omega,e_{\nu}\right]\right]+\cdots\frac{1}{m!}\left[\omega,\cdots\left[\omega,e_{\nu}\right]\right]+\cdots,\label{eq:23}
\end{equation}
using the commutation relations 
\begin{equation}
\left[J_{ab},\gamma_{c}\right]=\frac{1}{2}\left(\delta_{bc}\gamma_{a}-\delta_{ac}\gamma_{b}\right),\label{eq:24}
\end{equation}
we can see that the equation (\ref{eq:21}) can be rewritten as 
\begin{equation}
T_{\alpha\beta}^{a}\gamma_{a}=0.\label{eq:25}
\end{equation}
Since $T_{\alpha\beta}^{a}$ is antisymmetric in $\alpha$ and $\beta$,
the $\frac{1}{2}d^{2}(d-1)$ torsion conditions (\ref{eq:25}) in
$d$ dimensional space allow the same number of spin connections $\omega$$_{\alpha}^{ab}$
in cell $n$ to be fully expressed in terms of the soldering forms
in the same cell $n$ and the neighboring cells $n+1.$ For a given
spin connections, the affine connections can be found from (\ref{eq:15}).
Thus, we have shown that both affine and spin connections in a given
cell are completely determined by the soldering forms in neighboring
cells. Later, we will find the explicit solutions to the above equations
in two-dimensional space and discuss how to find the solution in closed
form for three- and four- dimensions space.

As we mentioned above, in discrete space the local symmetry group
with respect to which the theory must be invariant is the group of
rotations under which the vielbeins $\mathbf{v}_{a}(n)$ go to 
\begin{equation}
\widetilde{\mathbf{v}}_{a}(n)=\mathcal{R}_{a}^{b}(n)\mathbf{v}_{b}(n),\label{eq:26}
\end{equation}
where $\mathcal{R}_{a}^{b}(n)$ are the elements of the $SO(n)$ rotation
group in the vector representation. Moving to the spinor representations
we find that the element of the group of spin connections (\ref{eq:10})
under rotations transforms as 
\begin{equation}
\varOmega{}_{\beta}(n)\rightarrow\widetilde{\varOmega}_{\beta}(n)=\mathcal{R}(n)\varOmega{}_{\beta}(n)\mathcal{R}^{-1}(n+1_{\beta}),\label{eq:27}
\end{equation}
By introducing the trial function $f(n)$, we can rerwite this equation
as 
\begin{equation}
\widetilde{\varOmega}_{\beta}(n)f(n+1_{\beta})=\mathcal{R}(n)\varOmega{}_{\beta}(n)\mathcal{R}^{-1}(n+1_{\beta})f(n+1_{\beta}),\label{eq:28}
\end{equation}
or alternatively 
\begin{equation}
\widetilde{\varOmega}_{\beta}(n)E_{\beta}(n)f(n)=\mathcal{R}(n)\varOmega{}_{\beta}(n)E_{\beta}(n)\mathcal{R}^{-1}(n)f(n),\label{eq:29}
\end{equation}
from which we derive the covariant transformation law for the operators
$\varUpsilon{}_{\alpha}(n)$, introduced in (\ref{eq:22}), 
\begin{equation}
\varUpsilon{}_{\alpha}(n)\rightarrow\widetilde{\varUpsilon}_{\alpha}(n)=\mathcal{R}(n)\varUpsilon{}_{\alpha}(n)\mathcal{R}^{-1}(n).\label{eq:30}
\end{equation}
The next step is to define the curvature by considering a plaquette,
which starts in cell $n$ and extends to neighboring cells with the
operators $\varUpsilon{}_{\alpha}(n)$ and $\varUpsilon{}_{\beta}(n)$
and backwards 
\begin{align}
R_{\alpha\beta}(n) & =\frac{1}{2}\left(\varUpsilon{}_{\alpha}(n)\varUpsilon{}_{\beta}(n)\varUpsilon{}_{\alpha}^{-1}(n)\varUpsilon{}_{\beta}^{-1}(n)-\varUpsilon{}_{\beta}(n)\varUpsilon{}_{\alpha}(n)\varUpsilon{}_{\beta}^{-1}(n)\varUpsilon{}_{\alpha}^{-1}(n)\right)\label{eq:31}\\
 & =\frac{1}{2}\left(\varOmega{}_{\alpha}(n)\varOmega_{\beta}(n+1_{\alpha})\varOmega{}_{\alpha}^{-1}(n+1_{\beta})\varOmega_{\beta}^{-1}(n)-(\alpha\leftrightarrow\beta)\right)\label{eq:33}
\end{align}
The curvature is antisymmetric $R_{\alpha\beta}\left(n\right)=-R_{\beta\alpha}\left(n\right)$
and it is obviously covariant, i.e. under rotations $R_{\alpha\beta}\left(n\right)\rightarrow\widetilde{R}_{\alpha\beta}(n)=\mathcal{R}(n)R_{\alpha\beta}\left(n\right)\mathcal{R}^{-1}(n).$
The above definition agrees with the definition adopted for the Yang-Mills
curvature, except for the step of antisymmetrization in $\alpha\beta$
which is very important for our considerations. Our definition of
curvature corresponds to the difference that arises when a plaquette
is first circumnavigated counterclockwise and then clockwise. For
example, in three dimensions, for each cell there are three possible
plaquettes passing through adjacent cells. For $SO\left(2\right),$
$SO\left(3\right),$ $SO\left(4\right)$, one can prove that $R_{\alpha\beta}\left(n\right)$
become the elements of the Lie algebra of the corresponding group
and therefore, can be written as 
\begin{equation}
R_{\alpha\beta}\left(n\right)=R_{\alpha\beta}^{\quad cd}(n)J_{cd},\label{eq:34}
\end{equation}
where $J_{cd}$ are the generators of the rotation group in spinor
representation, defined in (\ref{eq:18}). By contracting indices
with the corresponding soldering forms, we can build the scalar curvature
as 
\begin{equation}
R(n)=R_{\alpha\beta}^{\quad cd}(n)\bar{e}_{c}^{\alpha}(n)\bar{e}_{d}^{\beta}(n).\label{eq:35}
\end{equation}
For the groups $SO\left(d\right),$ $d>4$ we project $R_{\alpha\beta}\left(n\right)$
on $J_{cd}$ by taking the trace to define 
\begin{equation}
R_{\alpha\beta}^{\quad cd}(n)=-2^{3-\left[\frac{d}{2}\right]}\mathrm{Tr}\left(R_{\alpha\beta}\left(n\right)J^{cd}\right).
\end{equation}

\section{Dirac action}

To generalize the Dirac equation for the case of discrete space, we
need to define the inner product for the spinors and determine the
hermitian operator which reproduces the well known Dirac equation
in the continuous limit. The inner product can be defined as 
\begin{equation}
\left(\psi,\psi\right)\equiv\sum_{n}\psi^{\dagger}\left(n\right)\psi\left(n\right),\label{eq:35b}
\end{equation}
while the natural candidate for Dirac operator is 
\begin{equation}
D(n)\equiv i\upsilon(n)\bar{e}^{\alpha}\left(n\right)\left(\varUpsilon_{\alpha}\left(n\right)-\varUpsilon_{\alpha}^{-1}\left(n\right)\right),\label{eq:35a}
\end{equation}
where $\bar{e}^{\alpha}\left(n\right)\equiv\bar{e}_{b}^{\alpha}\left(n\right)\gamma^{b}$
and the function $\upsilon(n)$ must still be determined by requirement
that $D$ is a hermitian operator, i.e. $\left(\psi,D\psi\right)=\left(D\psi,\psi\right)$.
Considering that 
\begin{equation}
D^{\dagger}(n)=i\left(\varUpsilon_{\alpha}\left(n\right)-\varUpsilon_{\alpha}^{-1}\left(n\right)\right)\upsilon(n)\bar{e}^{\alpha}\left(n\right),\label{eq:35c}
\end{equation}
and recalling the definition in (\ref{eq:22}), we conclude that $D^{\dagger}(n)=D(n)$
holds if the function $\upsilon(n)$ satisfies the following equation
\begin{equation}
\upsilon(n)\bar{e}^{\alpha}\left(n\right)\varOmega_{\alpha}(n)=\upsilon(n+1_{\alpha})\varOmega_{\alpha}(n)\bar{e}^{\alpha}\left(n+1_{\alpha}\right).\label{eq:35d}
\end{equation}
One can easily prove that this equation is invariant with respect
to the rotation group and thus $\upsilon(n)$ is invariant. As we
will show later, the solution for $\upsilon(n)$ in the continuous
limit becomes $\det(e_{\alpha}^{b})$. Thus, in the discrete case
the action for the Dirac spinors is 
\begin{equation}
S=\sum_{n}i\psi^{\dagger}\left(n\right)\upsilon(n)\bar{e}^{\alpha}\left(n\right)\left(\varUpsilon_{\alpha}\left(n\right)-\varUpsilon_{\alpha}^{-1}\left(n\right)\right)\psi\left(n\right).\label{eq:37}
\end{equation}
The gauge invariant action for discrete Euclidean gravity is accordingly
\begin{equation}
S=\sum_{n}\upsilon(n)R(n),\label{eq:36-1}
\end{equation}
where $\upsilon(n)$ is the solution of equation (\ref{eq:35d}).

\section{Example: two dimensional space}

Let us consider the simplest example of the two-dimensional discrete
space with the local $SO(2)$ tangent group and write down explicit
formulas for the spin connection and curvature in terms of the soldering
form. In the two-dimensional space, all cells are numbered by two
integers $n=(n^{1},n^{2}).$ The matrices satisfying the Clifford
algebra (\ref{eq:19}) can be taken as 
\begin{equation}
\gamma_{1}=\left(\begin{array}{cc}
1 & 0\\
0 & -1
\end{array}\right),\quad\gamma_{2}=\left(\begin{array}{cc}
0 & 1\\
1 & 0
\end{array}\right).\label{eq:38}
\end{equation}
and the single nonvanishing generator $J_{12}=-J_{21}$ , defined
in (\ref{eq:18}), becomes 
\begin{equation}
J_{12}=\frac{1}{4}\left(\begin{array}{cc}
0 & 1\\
-1 & 0
\end{array}\right)\equiv\frac{1}{4}\tau.\label{eq:39}
\end{equation}
Taking into account that 
\begin{equation}
\omega_{\alpha}=\omega_{\alpha}^{ab}(n^{1},n^{2})J_{ab}=\frac{1}{2}\omega_{\alpha}^{12}(n^{1},n^{2})\tau.\label{eq:40}
\end{equation}
we obtain the following expression for the spin connection elements
of the algebra (\ref{eq:10}) 
\begin{equation}
\varOmega_{\alpha}(n^{1},n^{2})=\cos\frac{1}{2}\omega_{\alpha}(n^{1},n^{2})+\tau\sin\frac{1}{2}\omega_{\alpha}(n^{1},n^{2}),\label{eq:41}
\end{equation}
where $\omega_{\alpha}(n^{1},n^{2})\equiv\omega_{\alpha}^{12}(n^{1},n^{2})$.
Using freedom in choice of gauge and partitioning the manifold into
cells (which in continuous limit corresponds to freedom in choice
of coordinates) we can set 
\begin{equation}
e_{1}^{1}=e_{2}^{2}=e\left(n^{1},n^{2}\right),\quad e_{2}^{1}=e_{1}^{2}=0.\label{eq:42}
\end{equation}
at each cell $n=(n^{1},n^{2})$. In this case, the torsion-free conditions
(\ref{eq:21}) simplify to 
\begin{align}
e(n^{1}+1,n^{2})\sin\omega_{1}(n^{1},n^{2})-e(n^{1},n^{2}+1)\cos\omega_{2}(n^{1},n^{2})+e(n^{1},n^{2}) & =0,\nonumber \\
e(n^{1}+1,n^{2})\cos\omega_{1}(n^{1},n^{2})+e(n^{1},n^{2}+1)\sin\omega_{2}(n^{1},n^{2})-e(n^{1},n^{2}) & =0.\label{eq:43}
\end{align}
Solving these equations gives the following explicit expressions for
$\omega_{\alpha}(n^{1},n^{2})$ in terms of soldering forms in three
adjacent cells 
\begin{align}
\omega_{1}(n^{1},n^{2}) & =\frac{\pi}{4}-\arcsin\left(\frac{e^{2}(n^{1}+1,n^{2})-e^{2}(n^{1},n^{2}+1)+2e^{2}(n^{1},n^{2})}{2\sqrt{2}e(n^{1}+1,n^{2})e(n^{1},n^{2})}\right),\nonumber \\
\omega_{2}(n^{1},n^{2}) & =\frac{\pi}{4}-\arccos\left(\frac{e^{2}(n^{1},n^{2}+1)-e^{2}(n^{1}+1,n^{2})+2e^{2}(n^{1},n^{2})}{2\sqrt{2}e(n^{1},n^{2}+1)e(n^{1},n^{2})}\right),\label{eq:44}
\end{align}
Substituting (\ref{eq:39})-(\ref{eq:41}) into (\ref{eq:33}) and
(\ref{eq:34}) we find that the only nonvanishing independent component
of the curvature is 
\begin{equation}
R_{12}^{\quad12}(n)=2\sin\left[\frac{1}{2}\left(\omega_{2}(n^{1}+1,n^{2})-\omega_{1}(n^{1},n^{2}+1)+\omega_{1}(n^{1},n^{2})-\omega_{2}(n^{1},n^{2})\right)\right],\label{eq:45}
\end{equation}
where $\omega_{1}$ and $\omega_{2}$ are given in (\ref{eq:44}).
The scalar curvature in this case is equal to 
\begin{equation}
R(n)=R_{\alpha\beta}^{\quad cd}(n)\bar{e}_{c}^{\alpha}(n)\bar{e}_{d}^{\beta}(n)=2R_{12}^{\quad12}(n)e^{-2}(n).\label{eq:46}
\end{equation}
Similar expressions for spin connection and curvature can be obtained
in three- and four- dimensional spaces, since $SO(3)$ and $SO(4)$
are locally isomorphic to $SU(2)$ and $SU(2)\times SU(2)$ groups,
respectively.

\section{Continuous limit}

So far, we have considered a discrete space with cells of elementary
volume set to one, and numbered the cells by ordered integers. To
obtain the continuous limit, we need to shrink the cells to the points.
If we introduce the variables 
\begin{equation}
x^{\alpha}=\epsilon n^{\alpha}.\label{eq:47-1}
\end{equation}
instead of integers, the volume of each cell vanishes when we take
the limit $\epsilon\rightarrow0$, and we expect to get the continuous
limit. However, one has to be careful. The situation is similar to
a straight line of length $L$ with $0<x<L$ divided by a series of
points with small spacing, and where the obtained cells are numbered
by integers $0<n<L/\epsilon$. It is obvious that the point with a
given $x_{0}$ moves to another cell $n$ as $\epsilon$ decreases.
Therefore, we must hold $x_{0}^{\alpha}$ characterizing the corresponding
point of the manifold in the continuous limit when $\epsilon\rightarrow0$.
This means that in this limit $n\rightarrow\infty$ for every single
point. For $\epsilon\neq1$ the shift operator (back to bold notation)
is defined as 
\begin{equation}
\mathbf{E}_{\alpha}(x)f(x)=f(x+\epsilon_{\alpha}),\label{eq:47a}
\end{equation}
where $x\equiv(x^{1},...,x^{d})$ and $\epsilon_{\alpha}\equiv\epsilon1_{\alpha}$
indicates which of $d$ coordinates has been changed, such as $\mathbf{E}_{2}(x)f(x)=f(x^{1},x^{2}+\epsilon_{2},...,x^{2})$.
We leave the index for $\epsilon$ in the lower position to avoid
the confusion that can arise from using Einstein's summation convention.
Accordingly, the tangent operators (\ref{eq:4c}) are acting as 
\begin{equation}
\mathbf{e_{\alpha}\left(\mathnormal{x}\right)\mathbf{=\mathnormal{\frac{1}{2\epsilon_{\alpha}}}\left(\mathbf{E}_{\alpha}\mathnormal{\left(x\right)}-\mathbf{E}_{\alpha}^{-1}\mathnormal{\left(x\right)}\right)\mathbf{\mathnormal{f(x)=\frac{f(x+\epsilon_{\alpha})-f(x-\epsilon_{\alpha})}{2\epsilon_{\alpha}}}.}}}\label{eq:47b}
\end{equation}
It follows that in the limit $\epsilon\rightarrow0$, 
\begin{equation}
\mathbf{e}_{\alpha}=\frac{\partial}{\partial x^{\alpha}},\label{eq:48b}
\end{equation}
i.e., in the continuous limit the tangent operators become the vectors
tangent to the corresponding coordinate lines. The formulas for parallel
transport (\ref{eq:8})-(\ref{eq:10}) in the case of $\epsilon\neq1$
are modified as, 
\begin{equation}
\mathbf{e}_{\alpha}^{p.t.}(x+\epsilon_{\beta}\rightarrow x)=\mathbf{e}_{\alpha}(x)+\Gamma_{\alpha\beta}^{\gamma}(x)\mathbf{e}_{\gamma}(x)\epsilon_{\beta}\label{eq:48c}
\end{equation}
and 
\begin{equation}
\mathbf{v}_{a}^{p.t.}(x+\epsilon_{\beta}\rightarrow x)=(\varOmega_{\beta}^{-1}(x))_{a}^{b}\mathbf{v}_{b}(x),\label{eq:48d}
\end{equation}
where $\varOmega_{\beta}^{-1}(x)$ is inverse to 
\begin{equation}
\varOmega{}_{\beta}(x)=\exp\left(\epsilon_{\beta}\omega_{\beta}^{cd}(x)J_{cd}\right).\label{eq:49a}
\end{equation}
We would like to emphasize again that no summation over $\beta$ is
assumed in (\ref{eq:48c}) and (\ref{eq:49a}) because these indices
have the same lower position, Taking the limit $\epsilon\rightarrow0$,
we obtain from these formulas the well-know expression for the covariant
derivatives 
\begin{equation}
\nabla_{\beta}\mathbf{e}_{\alpha}=\Gamma_{\alpha\beta}^{\gamma}\mathbf{e}_{\gamma},\qquad\nabla_{\beta}\mathbf{v}_{a}=\omega_{\beta a}^{\quad b}\mathbf{v}_{b}.\label{eq:50a}
\end{equation}
Next we find the solution of the equation (\ref{eq:35d}) for $\upsilon(x)$
in the limit $\epsilon\rightarrow0$. Expanding this equation in powers
of small $\epsilon$ and using the commutation relation (\ref{eq:24})
we find 
\begin{equation}
\left(\frac{\partial_{\alpha}\upsilon}{\upsilon}\bar{e}_{b}^{\alpha}+\partial_{\alpha}\bar{e}_{b}^{\alpha}+\omega_{\alpha b}^{c}\bar{e}_{c}^{\alpha}\right)\gamma^{b}\epsilon_{\alpha}+O\left(\epsilon^{2}\right)=0,\label{eq:51}
\end{equation}
where $\partial_{\alpha}\equiv\partial/\partial x^{\alpha}.$ At the
leading order, the expression in parenthesis should vanish for each
$b$. Multiplying it by $e_{\beta}^{b}$ and summing over $b$, we
get the following equation for $\upsilon$ in the limit $\epsilon\rightarrow0$,
\begin{equation}
\frac{\partial_{\beta}\upsilon}{\upsilon}+e_{\beta}^{b}\partial_{\alpha}\bar{e}_{b}^{\alpha}+\omega_{\alpha b}^{c}\bar{e}_{c}^{\alpha}e_{\beta}^{b}=0\label{eq:51a}
\end{equation}
To find the general solution of this equation, we note that at the
leading linear order in $\epsilon$, the no-torsion equation (\ref{eq:17})
becomes 
\begin{equation}
\partial_{\beta}e_{\alpha}^{a}-\omega_{\beta b}^{a}e_{\alpha}^{b}=\partial_{\alpha}e_{\beta}^{a}-\omega_{\alpha b}^{a}e_{\beta}^{b}.\label{51b}
\end{equation}
Multiplying this equation by $\bar{e}_{a}^{\alpha}$ and summing over
$\alpha$ and $a$ one obtains 
\begin{equation}
\frac{\partial_{\beta}\det(e_{\alpha}^{a})}{\det(e_{\alpha}^{a})}+e_{\beta}^{a}\partial_{\alpha}\bar{e}_{a}^{\alpha}+\omega_{\alpha b}^{a}\bar{e}_{a}^{\alpha}e_{\beta}^{b}=0,\label{eq:51d}
\end{equation}
and by comparing with (\ref{eq:51a}) we conclude that 
\begin{equation}
\upsilon=\det(e_{\alpha}^{a})\label{eq:51e}
\end{equation}
in the continuous limit.

The expression for the curvature (\ref{eq:33}) in the limit $\epsilon\rightarrow0$
becomes 
\begin{equation}
R_{\alpha\beta}(x)=\lim_{\epsilon\rightarrow0}\frac{1}{2\epsilon_{\alpha}\epsilon_{\beta}}\left(\varOmega{}_{\alpha}(x)\varOmega_{\beta}(x+\epsilon_{\alpha})\varOmega{}_{\alpha}^{-1}(x+\epsilon_{\beta})\varOmega_{\beta}^{-1}(x)-(\alpha\leftrightarrow\beta)\right).\label{eq:52}
\end{equation}
Substituting here (\ref{eq:49a}), computing the limit and projecting
$R_{\alpha\beta}$ on $J_{cd}$ (see, (\ref{eq:41})), we obtain the
following result for the components of the spin connection curvature
\begin{equation}
R_{\alpha\beta}^{\quad cd}(x)=\partial_{\alpha}\omega_{\beta}^{cd}-\partial_{\beta}\omega_{\alpha}^{cd}+\omega_{\alpha}^{cl}\omega_{\beta l}^{\quad d}-\omega_{\beta}^{cl}\omega_{\alpha l}^{\quad d},\label{eq:53}
\end{equation}
in full agreement with the known standard result.

Returning to the example of straight line, we now consider a function
$f(x)$. After discretizing the straight line, we assign to the function
$f(x)$ a value that it takes in the cell $n$, i.e. $f(n).$ It is
obvious that 
\begin{equation}
\lim_{\epsilon\rightarrow0}\sum_{n}\epsilon f(n)=\int f(x)dx.\label{eq:49}
\end{equation}
Therefore, for example, the action for gravity (\ref{eq:36-1}) in
the continuous limit becomes 
\begin{equation}
S=\int\det(e_{\alpha}^{b})R(x)dx^{1}...dx^{d}.\label{eq:50}
\end{equation}
Thus, we have proved that our theory of discrete manifolds reproduces
all results for the continuous manifold in the corresponding limit.

\section{Conclusions}

In summary, we have considered a space consisting of elementary cells
of a certain (e.g. Planck) volume and assumed that these cells have
no internal differentiable structure. Each of these cells is fully
characterized only by a finite number of operators and spin connections.
In such a discrete space, we defined the parallel transport and found
out how spin connections, torsion and curvature can be expressed in
terms of the soldering forms in the neighboring cells. The developed
theory of discrete space is explicitly invariant with respect to the
local rotation group. In our theory, the problem of failure of the
Liebnitz rule, which is usually an obstacle to the development of
the theory of discretized manifolds (see, e.g., \cite{Miller,Bochev})
is avoided since we use the spin connection group as a basis. Here
we are guided by the principle that the tangent group for spinors
is the rotation group $SO\left(d\right)$ which in the continuous
limit is connected to the base manifold through the soldering forms.
The freedom in the choice of elementary cells is reminiscent of diffeomorphism
invariance, which naturally reappears in the continuous limit. We
have shown that when the cells shrink to the points, we exactly reproduce
all the formulas for a differentiable manifold.

Unlike field theory, we need only to assign a finite number of degrees
of freedom to each cell for each field. This could lead to the appearance
of a natural ultraviolet cut-off in the field theory, a problem that
requires further investigation. The developed mathematical structure
is well suited to describe the quanta of geometry in the noncommutative
approach \cite{CCM} and quantized black holes \cite{BM}.

Following canonical quantum gravity, it is natural to assume that
only the three-dimensional space-like hyperspaces have a structure
of discrete cells, the quanta of geometry. This is supported by the
formulation of the Cauchy problem in general relativity and serves
as an explicit manifestation of the idea of a finite number of degrees
of freedom per Planck volume. In this case, in the expanding universe,
the number of elementary quanta and hence the number of degrees of
freedom, increases with time. This leads to a rather natural picture
of emergent space, where the quanta of geometry either emerge or disappear
with time. For example, one could start with a single elementary quantum
and generate an arbitrarily large number of quanta as a result of
inflationary expansion. In future publications, we will show how the
formalism developed in this work provides a simple and adequate way
to describe this picture. Having constructed the discrete analog
of the Einstein action for discrete spaces, there are a large number
of applications of our formalism to gravity. In particular, for the
dimensions $d=2,3,4,$ the components of the curvature tensor can
be calculated in closed form at any point.

\textbf{\large{}{}{}{}{}{Acknowledgments}}{\large\par}

The work of A. H. C is supported in part by the National Science Foundation
Grant No. Phys-1912998 and by the Humboldt Foundation. The work of
V.M. is supported by the Deutsche Forschungsgemeinschaft (DFG, German
Research Foundation) under Germany's Excellence Strategy -- EXC-2111
-- 390814868.

\end{document}